\documentclass[aps,pre,reprint,showpacs,superscriptaddress]{revtex4-2}

\usepackage{graphicx}
\usepackage{dcolumn}
\usepackage{bm}
\usepackage{xcolor}  
\expandafter\let\csname equation*\endcsname\relax
\expandafter\let\csname endequation*\endcsname\relax
\usepackage{amsmath}
\usepackage{graphicx}
\usepackage{soul}
\usepackage{physics}
\emergencystretch=\maxdimen
\usepackage[export]{adjustbox}
\begin{document}

\title{Thermodynamic interpretation to Stochastic Fisher Information and Single-Trajectory Speed Limits}

\author{Pedro \surname{B.~Melo}}
\email{pedrobmelo@aluno.puc-rio.br}
 \affiliation{Departamento de F\'isica, Pontif\'icia Universidade Cat\'olica, 22452-970, Rio de Janeiro RJ, Brazil}
 \author{Fernando \surname{Iemini}}
 \affiliation{Instituto de F\'isica, Universidade Federal Fluminense, 24210-346, Niter\'oi RJ, Brazil}
 \author{Diogo \surname{O.~Soares-Pinto}}
 \affiliation{Instituto de Física de São Carlos, Universidade de São Paulo, 13560-970 São Carlos SP, Brazil}
 \author{S\'ilvio M. \surname{Duarte~Queir\'os}}
 \affiliation{Centro Brasileiro de Pesquisas F\'isicas, 22290-180, Rio de Janeiro RJ, Brazil}
 \affiliation{INCT-Sistemas Complexos, Brazil}
 \author{Welles \surname{A.~M.~Morgado}}
\affiliation{Departamento de F\'isica, Pontif\'icia Universidade Cat\'olica, 22452-970, Rio de Janeiro RJ, Brazil}%
\affiliation{INCT-Sistemas Complexos, Brazil}
\date{\today}

\begin{abstract}
 The Fisher information (FI) metric is a Riemannian metric that allows a geometric treatment of stochastic thermodynamics, introducing the possibility of computing thermodynamic lengths and deviations from equilibrium. At the trajectory level, a related quantity can be introduced, the stochastic Fisher information (SFI), which on average, is equivalent to the FI. In this work, we discuss two fundamental questions regarding the SFI; namely, (i) what is the thermodynamic interpretation to the SFI, and (ii) are there any trajectory-level thermodynamic bounds . We find that, contrary to previous results in the literature for the FI, the thermodynamic interpretation of the SFI depends only on the entropy produced by the system and on the thermodynamic force. Moreover, we find that the SFI allows one to derive single-trajectory speed limits, which we demonstrate to hold for a Brownian particle under a saturating drive force and a Brownian particle under a decreasing drive force. From the ensemble of single-trajectory bounds, one can derive a hierarchy of average speed limits that are always less tight than the one derived from the FI. We test our results for speed limits on the adopted models and find that the hierarchy of average speed limits is respected and that the single-trajectory speed limits behave qualitatively similar to the average and stochastic speed limits, with some trajectories achieving velocities higher than the tightest average bound whenever it does not saturate. Our results open avenues for the exploration of uncertainty relations at the trajectory level. 
 \end{abstract}
\maketitle


\section{Introduction} 

The roots of Stochastic Thermodynamics stem from the introduction of fluctuation theorems (FTs) \cite{Evans_Cohen1993, Gallavotti1995, GallavottiCohen1995, jarzynski1997, Crooks1999, martins2025briefintroductionfluctuationtheorems}. In essence, FTs show that the laws of thermodynamics for non-equilibrium regimes are probabilistic instead of deterministic, even though the first law appears deterministic in such regimes. To assess the physics of the systems of interest, including heat, work, and entropy \cite{Sevick2008,Seifert_2012, Mauro2021, martins2025briefintroductionfluctuationtheorems}, one needs to compute higher order moments than just the average. This probabilistic framework is called Stochastic Thermodynamics and in this context one can also formulate a set of laws analogous to the equilibrium thermodynamic laws~\cite{sekimoto2010,peliti2021}.  For the second law, it becomes possible to calculate stochastic entropy and stochastic entropy production along a single trajectory \cite{seifert2005, Mauro2021}, with its probability distributions following a fluctuation theorem.~These results are experimentally accessible for systems such as optical tweezers \cite{Collin2005, ciliberto2017} or even in the quantum regime \cite{Batalhao2014, Micadei2019}.

\begin{figure}[ht!]
    \centering
    \includegraphics[width=1.0\linewidth]{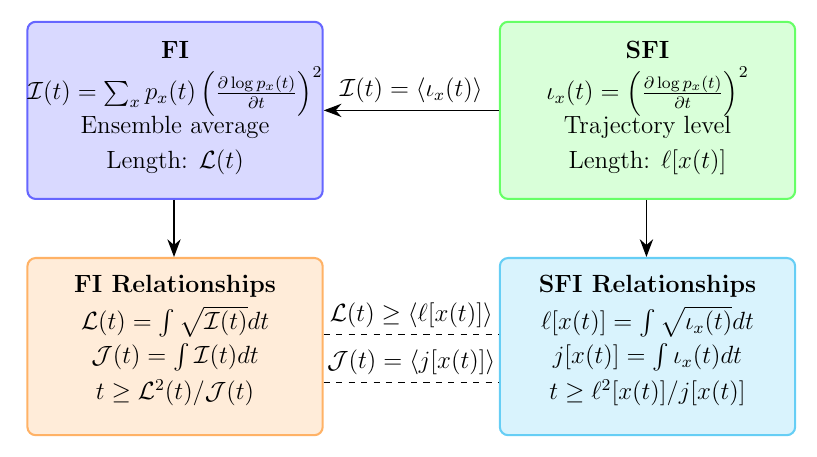}
    \caption{Schematic representation of the similarities between the SFI taking $\theta = t$. The upper boxes indicate the FI and the SFI frameworks, with their corresponding definitions along, and the symbol that represents the respective length in each framework. The arrow between upper balloons indicates that the FI can be obtained by averaging the SFI. The lower boxes represent the geometric relationships for both FI and SFI, for the respective lengths, action along the curve for each metric, and each respective uncertainty relation. The dashed lines between the lower ballons indicate the relations between the thermodynamic length $\mathcal{L}(t)$ and the average stochastic length $\langle l[x(t)]\rangle$, as well as between the thermodynamic action $\mathcal{J}(t)$ and the average stochastic action $\langle j[x(t)]\rangle$.}
    \label{fig:enter-label}
\end{figure}

A geometrical approach to thermodynamics is also meaningful and has first been established for a slow driving regime in usual thermodynamics \cite{Weinhold1975, Ruppeiner1979, Salamon1983, Brody1995, Ruppeiner1995}. More recently, an approach valid for more general non-equilibrium regimes have been established, and extended to stochastic systems as well~\cite{crooks2007, Crooks2009, Sivak2012, Ito2018, ito2020}. Central to this approach is information geometry, a framework that introduces statistical manifolds, \textit{i.e.} Riemannian manifolds with probability distributions as points in state space \cite{amari2000methods}. By the \v{C}encov's theorem one learns that the only metric for the probability manifold that is contractive under noisy transformations is the Fisher information (FI) metric~\cite{fisher1922, Fisher1925}. The FI is a  quantity that has been multidisciplinarily employed in areas such as quantum information \cite{Escher2011,Toth2012,Pires2016,Liu_2020, min2022, Yadin2023, Iemini2024, Bettmann2025}, ecology \cite{Mayer2006}, and finance \cite{sahalia2008}, to mention a few. This quantity can be interpreted as a way to measure the amount of information a random variable $X$ has about an unknown parameter $\theta$. This interpretation is synthesized in the form of the Cram\'er-Rao (CR) bound \cite{Rao1992}. It states that for an unknown parameter $\theta$, with an estimator $\hat{\Theta}(\theta)$ (for instance, a sample mean of a probability distribution of $\theta$ within a data set), the precision of $\hat{\Theta}(\theta)$ is at most the FI. 

The FI and the CR bound are directly related to thermodynamic inequalities, such as the thermodynamic uncertainty relations (TURs) \cite{barato2015, Salazar2022, TanVanVu2023}, the kinetic uncertainty relations (KURs) \cite{Dechant2020}, and the stochastic speed limits \cite{Shiraishi2018,Gupta2020,TanVanVu2020}. They relate a signal-to-noise ratio (in the case of TURs) and the speed of transformation of states (in the case of speed limits) to the entropy production rate. In the quantum regime, there is known to exist a quantum FI that measures the amount of information a quantum state has about an unknown parameter $\theta$. The QFI is also directly related to (quantum) speed limits \cite{Taddei2013, delCampo2013, Pires2016, Deffner_2017, nishiyama2025}, that roughly speaking are derived from time-energy uncertainty relation. The stochastic speed limits (obtained by FI metrics) have an analogous interpretation to quantum speed limits \cite{Okuyama2018, Shanahan2018}. Near-equilibrium, the stochastic speed limits affirm that the time taken for a system to go from an initial state to a final state is lower bounded by the inverse of variance of energy \cite{Ito2018}. 

Recently, a stochastic quantity analogous to FI was introduced, dubbed stochastic Fisher information (SFI). For a parameter $\theta$, the SFI is a random variable with a probability distribution whose average reproduces the FI of the same $\theta$ \cite{melo2024stochasticthermodynamicsfisherinformation}. The SFI distribution was also discovered to be subject to fluctuation theorems \cite{Melo_2025}. As a metric, the length given by the SFI is called stochastic length $\ell[x(t)]$, and is dependent on the trajectory followed between two states, analogous to the length given by the FI, called thermodynamic length $\mathcal{L}(t)$ \cite{crooks2007}. Even though the FI is univocally recovered by the SFI, the length $\ell[x(t)]$ does not recover univocally the thermodynamic length $\mathcal{L}(t)$. Instead, the average $\ell[x(t)]$ follows an uncertainty relation with $\mathcal{L}(t)$. 

In the SFI setting, another quantities of interest can also be defined in the same fashion as metric-related quantities for the FI. One of such is the stochastic action of the curve $j[x(t)]$, which receives such name because it is analogous to the kinect energy \cite{crooks2007}. The mean stochastic action is equivalent to the thermodynamic action $\mathcal{J}(t)$. To appoint the definitions and similarities between the SFI and the FI, we show in Fig. \ref{fig:enter-label} a schematic representation of both frameworks for the parameter $\theta = t$.

In this work, we shed light onto two cornerstone questions about the SFI with time as the unknown parameter and its stochastic geometric framework. We found that the thermodynamic interpretation of the SFI cannot be expressed as the rates of thermodynamic force and of entropy produced by the heat bath, contrary to the results for the FI in the literature \cite{Ito2018}. Our results show that the thermodynamic intepretation to the SFI is entirely related to the thermodynamic force (not its rate in time) and the rate of system's entropy production rate. The second question is about the possibility of single-trajectory speed limits due to the SFI, much like the stochastic speed limits for the FI \cite{Shiraishi2018}. We found that one can derive single-trajectory speed limits, that on trajectory level can become tighter than the stochastic speed limit, apart from intervals where this bound saturates. Our adopted examples also show that the single-trajectory bound behaves qualitatively similar to the average bounds for all trajectories.

We organize this paper as follows: In section II, we generalize the SFI framework for discrete energy level systems. In section III, we derive a stochastic thermodynamic interpretation of the SFI. In section IV, we derive the single-trajectory speed limits and two new average speed limits. To test our results of single-trajectory speed limit, we adopt the examples of a Brownian particle under a saturating drive force, and under a decreasing drive force in section V. Finally, we make our conclusions in section VI.

\section{Generalization of SFI for systems of discrete energy levels}

To broaden the interpretation of the recent results on stochastic Fisher information \cite{melo2024classicalgeometricfluctuationrelations}, consider the probability distribution of the position of a Brownian particle at a time $t$ (position means spatial position for the continuous case, or the energy level for the discrete case) $P(X) = \{p_{i}(t)\}$, such that $0\le p_{i}(t) \le 1$ for $i = \{1,\dots,N\}$, and $\sum_{i=1}^{N} p_{i}(t) = 1$, for all $t$. In this section, let us consider the discrete case. The master equation has a transition rate matrix, ${\bf{W}}(t)$, the elements of which, $W_{ij}(t)$, indicate a transition rate from state $i$ to state $j$ at time $t$. The master equation that describes the time evolution is given by,
\begin{equation}
    \dv{p_{i}(t)}{t} = \sum_{j \ne i}(W_{ij}(t) p_{j}(t) - W_{ji}(t)p_{i}(t)),\label{eq:Master_equation}
\end{equation}
where the first term of the right side indicates the influx from $j \ne i$ states to $i$ state, and the second term the outflux from $i$ to $j$ states. $\bf{W}$ satisfies normalization condition $\sum_{i}W_{ij} = 0$ and non-negativity $W_{ij} > 0$ for $i \ne j$. 

Assume the set of $M$ control parameters $\theta(t) = \{\theta_{1}(t), \dots,\theta_{M}(t)\}$ for which the system depends on, and that define the average ($\textit{i.e.}$ deterministic) path taken by the system. The coordinates of a statistical manifold depend on $\lambda(t)$, with its points being given by $\{p(x,t|\theta(t))\}$. The appropriate metric that equips such manifold with a distance measure is the Fisher information (FI) matrix, $g_{ij}$. It means that for a measure of distance ${\rm d}s^2 = \frac{1}{4}\sum_{i,j} g_{ij} {\rm d}\theta_{i}{\rm d}\theta_{j}$ we have
\begin{equation}
    g_{ij}(\theta) = \sum_{x} p_{x}(\theta)\pdv{\log p_{x}(\theta)}{\theta_{i}}\pdv{\log p_{x}(\theta)}{\theta_{j}}.
\end{equation}

We simplify the problem by taking the only parameter for the FI as time $\theta(t) = t$. It means we have
\begin{equation}
    \mathcal{I}(t) \equiv \sum_{ij}\dv{\theta_{i}}{t} g_{ij}(\theta) \dv{\theta_{j}}{t} = \sum_{i} p_{i}(t) \left( \pdv{\log p_{i}(t)}{t}\right)^2,
\end{equation}
which allows us to rewrite the measure of distance as ${\rm d}m^2 =  \mathcal{I}(t){\rm d}t^2$. The length over a given path $\gamma$ is given by 
\begin{equation}
    \mathcal{L}(t) \equiv \int_{\gamma} {\rm d}m = \int_{0}^{t}{\rm d}t'\sqrt{\mathcal{I}(t')}.
\end{equation}

An alternative, analogous metric of distance for stochastic statistical manifolds \cite{Höhle2001} is the stochastic Fisher information \cite{melo2024stochasticthermodynamicsfisherinformation} that, with respect to time, reads
\begin{equation}
    \iota_{x}(t) = \left(\pdv{\log p_{x}(t)}{t}\right)^2,
    \label{eq:sfi_def}
\end{equation}
such that a measure of stochastic distance becomes ${\rm d}s^2(x,t) = \iota_{x}(t) {\rm d}t^2$. The length over a stochastic path $\gamma[x(t)]$, called stochastic length, is given by
\begin{equation}
    \ell[x(t)] \equiv \int_{\gamma[x(t)]} {\rm d}s(x,t) = \int_{0}^{t}{\rm d}t' \sqrt{\iota_{x}(t')}. 
\end{equation}

The quantity $\iota_{x}(t)$ is a stochastic variable that follows a probability distribution which on average is equivalent to the FI. Let us now consider a process between two equilibrium states where the system is taken arbitrarily out of equilibrium due to a protocol $\lambda (t)$. Due to the irreversible character of the physical process, forward and backward protocols will generate fundamentally different variations of entropy, with nonvanishing net entropy production between both processes. Assuming $P_{F}(\iota,t)$ as the distribution of $\iota_{x}(t)$ for the ensemble of forward trajectories and $P_{B}(\iota,t)$ as the distribution $\iota_{x}(t)$ for the ensemble of backward trajectories, the detailed geometric fluctuation relation states that
\begin{equation}
    \frac{P_{F}(\iota; t)}{P_{B}(\iota;t)} = \exp \{\beta q[x(t)] + \ell[x(t)]\},
\end{equation}
where $\beta \equiv \frac{1}{k_{B}T}$, and $q[x(t)]$ is the heat that flows from the heat bath to the system along the trajectory defined by $[x(t)]$. 

The integral geometric fluctuation relation states that 
\begin{equation}
    \langle \exp\{\beta q[x(t)] + \ell[x(t)]\}\rangle = 1, 
\end{equation}
or via the Jensen relation 
\begin{equation}
    \langle \ell[x(t)] + \beta q[x(t)]\rangle \ge 0,
    \label{eq:JI_integralFR}
\end{equation}
meaning that the average stochastic length is lower bounded by the entropy that outflows from the bath to the system. To give meaning to Eq.~(\ref{eq:JI_integralFR}), it is useful to recover weaker inequalities in terms of previously known physical observables. Restating the definition of $\mathcal{L}(t)$ in terms of the SFI yields
\begin{equation}
    \mathcal{L}(t) = \int~{\rm d}t'\sqrt{\langle \iota_{x}(t')\rangle},
\end{equation}
from which assuming the Cauchy-Schwarz inequality we retrieve $\mathcal{L}(t)\ge \langle \ell[x(t)]\rangle$. Plugging this inequality into Eq.~(\ref{eq:JI_integralFR}) it gives
\begin{equation}
        \mathcal{L}(t) \ge  \langle\ell[x(t)] \rangle \ge -\beta\langle q[x(t)]\rangle .
\end{equation}
It means that the thermodynamic length is part of a (weaker) version of the integral geometric FR, $\mathcal{L}(t) + \beta\langle q[x(t)]\rangle \ge 0$. This inequality is straightforward, since it holds for all cases, such as $\mathcal{L}(t) \ge 0$ and $\beta \langle q[x(t)]\rangle \ge 0$. Therefore, this bound is saturated only for equilibrium regimes. 

Contrary to the $S_{2}$ manifold described by a Fisher metric that gives a sphere surface of radius $2$ \cite{Ito2018}, the manifold described by the stochastic Fisher metric fluctuates in time.~It means that there is not a well-defined expression for the shortest length between two distributions ${\textbf{p}}_{{\rm ini}}$ and ${\textbf{p}}_{{\rm fin}}$.~Instead, the minimal stochastic length can be considered as
\begin{eqnarray}
    \ell_{{\rm min}}[x(t)]= \inf_{\gamma} \ell[x(t)],
\end{eqnarray}
$\gamma$ represents the ensemble of stochastic lengths.



\section{Stochastic Thermodynamics interpretation of SFI \label{sec:III}}

To enrich our discussion and provide a more complete picture, we follow recent results \cite{Ito2018,Nicholson2018} that give a thermodynamic interpretation to information geometry in classical stochastic systems with a dynamics described by Markovian master equations, namely the Schnakenberg network theory \cite{Schnakenberg1976}. Our objective is to propose an interpretation to the case of a stochastic metric.

Noticing that one can define the dynamics described by the master equation in Eq.~(\ref{eq:Master_equation}) for a system with $N$ discrete states, in terms of probability currents $J_{x'\rightarrow x} (t) = (W_{x'\rightarrow x}(t) p_{x'}(t) - W_{x \rightarrow x'}(t)p_{x}(t))$, we have
\begin{equation}
    \dv{p_{x}(t)}{t} =\sum_{x' \ne x} J_{x'\rightarrow x}(t),
\end{equation}
where $J_{x' \rightarrow x} (t) = 0$ for a reversible dynamics of equilibrium, and $W_{x\rightarrow x}(t) = -\sum_{x'}W_{x\rightarrow x'}(t)$ for probability conservation. 

We impose local detailed balance, meaning the variation of stochastic entropy of the system $s_{x(t)}^{\rm s} (t)$ along a trajectory $\{x(t)\}$ and environment $\Delta s_{x' \rightarrow x}^{\rm e}(t)$ are given by
\begin{eqnarray}
   s_{x(t)}^{\rm s} (t) = -\log p_{x(t)}(t),~~~~~~\\
   \Delta s^{\rm e}_{x' \rightarrow x}(t) = -\log \frac{W_{x'\rightarrow x}(t)}{W_{x\rightarrow x'}(t)},&&  
\end{eqnarray}
where the system entropy is also known as surprisal and the variation of system entropy between two states is also known as relative surprisal. We also introduce the conjugated thermodynamic force \cite{Esposito2010}%
\begin{align}
    F_{x \rightarrow x'}(t) &\equiv -\log \frac{W_{x\rightarrow x'}(t)p_{x}(t)}{W_{x'\rightarrow x}(t)p_{x'}(t)},\\
    &= -\log \frac{W_{x\rightarrow x'}(t)}{W_{x'\rightarrow x}(t)} - \log \frac{p_{x}(t)}{p_{x'}(t)},\\
    &= \Delta s_{x\rightarrow x'}^{{\rm e}}(t) + \Delta s_{x\rightarrow x'}^{{\rm s}}(t) ,
\end{align}
where $\Delta s_{x\rightarrow x'}^{\rm{s}}(t) \equiv -\log(p_{x}(t)/p_{x'}(t))$ represents the system stochastic entropy change as between $x$ and $x'$ states. Using the thermodynamic force one can rewrite Eq.~(\ref{eq:Master_equation}) as 
\begin{equation}
    \dv{p_{x}(t)}{t} = \sum_{x' = 0}^{N} W_{x\rightarrow x'}p_{x}(t)e^{-F_{x\rightarrow x'}}.
    \label{eq:Master_eq_force}
\end{equation}

We now proceed to rewrite the relation between the observables of stochastic thermodynamics and the stochastic line element.~Keeping in mind that the distance ${\rm d}s^2$ can be represented as an average of the measure of stochastic distance, \textit{i.e.}, ${\rm d}s^2 = \mathcal{I}(t){\rm d}t^2 = \langle\iota_{x}(t)\rangle{\rm d}t^2$ and that the time SFI is a time-local quantity, meaning $x$ is considered as a time-independent quantity and $p(x,t)$ effectively represents the marginal probability distribution at $p(x,t) = \langle \delta(x(t) - t)\rangle$, we follow the description of Ref. \cite{Ito2018} where
\begin{equation}
    \frac{{\rm d}m^2}{{\rm d}t^2} = \sum_{x} p_{x}(t) \dv{t}\left(-\frac{1}{p_{x}(t)}\dv{p_{x}(t)}{t}\right)\label{eq:av_line_elemnt},
\end{equation}
and, without the use of averages defined by the weighted sum of $p_{x}(t)$, we have for the stochastic metric
\begin{align}
    \frac{{\rm d}s^2(x,t)}{{\rm d}t^2} &= \left(\dv{\log p_{x}(t)}{t}\right)^2 \label{eq:19}\\&=  \frac{1}{p_{x}(t)}\dv{t}\left(\dv{p_{x}(t)}{t}\right) - \dv{t} \left( \dv{\log p_{x}(t)}{t}\right),\label{eq:20}\\
    &=\left(\sum_{x'} W_{x\rightarrow x'}(t)e^{-F_{x\rightarrow x'}}\dv{\log p_{x}(t)}{t}\right).\label{eq:21}
\end{align}
The time derivative of $\log p_{x}(t)$ in Eq.~(\ref{eq:21}) can be understood as a partial time derivative for $x \rightarrow x(t)$. This allows one to rewrite Eq.~(\ref{eq:21}) as
\begin{align}
\left.\frac{{\rm d}s^2(x,t)}{{\rm d}t^2}\right|_{x(t)} &=\left(\sum_{x'} W_{x\rightarrow x'}(t)e^{-F_{x\rightarrow x'}}\left.\pdv{\log p_{x}(t)}{t}\right|_{x(t)}\right),\label{Eq:24}
\end{align}
which is precisely the explicit contribution of the system entropy rate $\dot{s}_{x(t)}^{\rm{s}}(t)$ \cite{seifert2005}. It means that one is able to rewrite Eq.~(\ref{Eq:24}) as
\begin{align}
    \frac{{\rm d}s^2(x,t)}{{\rm d}t^2} &=\left(\sum_{x'} W_{x\rightarrow x'}(t)e^{-F_{x\rightarrow x'}(t)}\left(-\dot{s}_{tot}(x,t) \nonumber\right. \right.\\&\left. \left. -\sum_{j}\delta(t - t_j)\log\frac{p_{x^{+}}(t_j)W_{x^+\rightarrow x^-}(t_j)}{p_{x^-}(t_j)W_{x^- \rightarrow x^+}(t_j)}\right)\right),\\
    &= \left(\overline{e^{-F_{x\rightarrow x'}}\left(-\dot{s}_{tot} +\sum_{j}\delta(t - t_j)F_{x_{j}^{-}\rightarrow x_{j}^{+}}\right)}\right),\label{eq:28}
\end{align}
where $\dot{s}_{tot} \equiv \dot{s}_{tot}(x(t),t) = \dot{s}_{x(t)}^{{\rm s}}(t) + \dot{s}_{x'\rightarrow x}^{\rm{e}} (t)$ represents the rate of total entropy production \cite{seifert2005}. Here, $\overline{A_{x'\rightarrow x}} \equiv \sum_{x'}W_{x'\rightarrow x} A_{x'\rightarrow x}$ is the rate-weighted expected value of $A_{x'\rightarrow x}$. These results are detailed in Appendix \ref{app:1}. The above result differs from the ones for conventional FI as in Ref.~\cite{Ito2018}, even though on average they remain equivalent. This happens because we do not take the average contribution of the line element. Equation~(\ref{eq:28}) is an expansion of the original SFI in terms of a corresponding term to the line element of FI as in Eq.~(\ref{eq:av_line_elemnt}). Even though the SFI is local in time, and therefore in principle does not consider the whole $x(t)$ trajectory, one needs to assume the limit $x \rightarrow x(t)$ to compute length-related quantities, thus rendering our derivation meaningful.

To expand our physical insight about the SFI, take for example Eq.~(\ref{eq:sfi_def}). In the canonical ensemble, with the probability distribution of the $x$-th level being given by $p_{x}(t) = e^{-\beta \varepsilon_{x}(t)}/Z(\beta,t)$, where $Z(\beta,t) = \sum_{x'}e^{-\beta \varepsilon_{x'}(t)}$, the stochastic line element becomes 
\begin{align}
\frac{{\rm d}s^2(x,t)}{{\rm d}t^2} = \beta^2\left(\dv{\varepsilon_{x}(t)}{t} + \dv{\mathcal{F}(t)}{t}\right)^2,
\end{align}
with $\mathcal{F}(t) = - \beta^{-1}\log Z(\beta,t)$ being the free energy at time $t$. By substituting $\mathcal{F}(t)$, we get that
\begin{equation}
    \frac{{\rm d}s^2(x,t)}{{\rm d}t^2} = \beta^2\left(\dv{\varepsilon_{x}(t)}{t} - \left\langle \dv{\varepsilon_{x}(t)}{t} \right\rangle\right)^2,
\end{equation}
which represents the contributions to the variance of the rate of change of the energy, which on average agree to previous results in the literature \cite{Crooks2009}. 

The results of Eq.~(\ref{eq:28}) can also be drawn from an overdamped Langevin perspective \cite{vanKampen}. A Brownian overdamped particle has a position probability distribution $P(x,t)$ where $x \equiv x(t)$ is the trajectory. The correspondent Fokker-Planck equation that describes the dynamics of $P(x,t)$ is given by
\begin{align}
    \pdv{P(x,t)}{t} &= -\pdv{x}J(x,t)\nonumber\\&= -\pdv{x}\left(f(x,t) - D\pdv{x}\right)P(x,t),
\end{align}
with $J(x,t)$ representing the probability flux, $f(x,t)$ representing the conservative and dissipative forces of the system, and $D$ the diffusion coefficient. Taking $P(x,t)$ locally in time, it becomes possible to restate results from Eq.~(\ref{eq:28}) as \cite{seifert2005}
\begin{align}
    \left.\frac{{\rm d}s^2(x,t)}{{\rm d}t^2}\right|_{x(t)} = \left(-\dot{s}_{\rm tot} + \left.\frac{J^2(x,t)}{D~P(x,t)}\right|_{x(t)}\dot{x}(t)\right)^2, 
\end{align}
($\dot{s}_{{\rm tot}} \equiv \left.\dot{s}_{{\rm tot}}(x,t)\right|_{x(t)}$). To obtain this result one must change $(\partial_{t} \log P(x,t))_{x(t)} $ by the combination of $\dot{s}_{{\rm tot}}$ and $(J^2/(D P(x,t)))_{x(t)} \dot{x}(t)$ \cite{seifert2005}. The above result is equivalent to Eq.~(\ref{eq:28}).

In the following section, we deal with another facet of the SFI. From the definition of stochastic length, it becomes possible to derive speed limits on trajectory level that follow relations similar to the known previous results of the literature for FI and thermodynamic length.

\section{Single-trajectory stochastic speed limits}

For a path in a statistical manifold, given that the Fisher information is a metric relating a line element between two distributions, the stochastic Fisher information relates a stochastic version of the line element,
\begin{equation}
    {\rm d}\ell^2(x,t) = \iota_{x}(t)~{\rm d} t^2, 
\end{equation}
analogously to the relation defined for the conventional Fisher information metric. Taking into account this metric, there exists a geodesic that represents the smallest distance between two probabilities, $p(x_{\text{ini}})$ and $p(x_{\text{fin}})$. The deviation from the geodesic is measured by the thermodynamic action $\mathcal{J}(t)$, which is given by
\begin{equation}
    \mathcal{J} (t)= \int_{0}^{t}{\rm d}t \left(\frac{{\rm d}s^2}{{\rm d}t^2}\right) = \int_{0}^{t}{\rm d}t~ \mathcal{I}(t).
\end{equation}
$\mathcal{J}(t)$ has such name because of its resemblance to the kinect integral energy. Moreover, $\delta(t) = \Delta t\mathcal{J}(t)$ is taken as a thermodynamic divergence from the geodesic path.

On SFI framework, the equivalent to the action $\mathcal{J}$ is the stochastic action $j[x(t)]$, written as
\begin{equation}
j[x(t)] \equiv \int_{0}^{t}dt~\iota_{x}(t), 
\end{equation}
where $\mathcal{J} = \langle j[x(t)]\rangle$. For equilibrium distributions, $j[x(t)]$ reduces to
\begin{equation}
    j[x(t)] = \int {\rm d}t~\beta^2\left(\frac{{\rm d}\varepsilon_{x}(t)}{{\rm d}t} - \left\langle\frac{{\rm d}\varepsilon_{x}(t)}{{\rm d}t}\right\rangle\right)^2.
\end{equation}

As in Ref.~\cite{Ito2018}, by making use of the Cauchy-Schwarz inequality $\int {\rm d}t ^\prime \int{\rm d}t[{\rm d}s^2(x,t)/{\rm d}t^2] \ge [\int {\rm d}t ({\rm d}s(x,t)/{\rm d}t)]^2$ for a square stochastic length $\ell^{2}[x(t)]$, we get
\begin{equation}
   \Delta t \ge \frac{\ell^2[x(t)]}{j[x(t)]}, 
    \label{eq:inequality_action}
\end{equation}
which represent a time-cost uncertainty relation, given we have a stochastic metric. Again, near equilibrium we have a term that is proportional to the variation of dynamically averaged bath entropy that flows from system to heat bath. Equation~(\ref{eq:inequality_action}) is analogous to the speed limit for thermodynamic lengths, given by
\begin{equation}
    \Delta t \ge \frac{\mathcal{L}^2(t)}{\mathcal{J}(t)}.
    \label{eq:FI_speed_limits}
\end{equation}

For the shortest stochastic path, another inequality emerges, meaning
\begin{equation}
    \Delta t \ge \frac{\ell^2[x(t)]}{j[x(t)]} \ge \frac{\ell^{2}_{{\rm{min}}}[x(t)]}{j[x(t)]}
\end{equation}

Following \cite{Sivak2012} we have that $\mathcal{J}(t)$, the average stochastic action, is precisely equivalent to $ \langle j[x(t)]\rangle \equiv W_{irr}(t)$, where $W_{irr}(t)$ is the irreversible work produced for a protocol $\lambda(t)$, from $\lambda(0)$ to $\lambda(t)$ near-equilibrium. $W_{irr}(t)$ can be defined as the work realized by an external agent that does not turn into system's free energy. 

This equivalence allows the existence of a physical intuition behind Eq.~(\ref{eq:FI_speed_limits}). Considering Eq.~(\ref{eq:FI_speed_limits}), we have $W_{irr}(t) \gtrsim \mathcal{L}(t)/\Delta t$, which means that the velocity of the system transformation between two states is limited by a quantity proportional to the entropy produced during such transformation. For Eq.~(\ref{eq:inequality_action}), the physical intuition remains similar, with $j[x(t)]$ representing the contributions of the stochastic irreversible work $w_{irr}[x(t)]$ along a trajectory, and the entropic velocity is also stochastic.

To derive other physical observables, we calculate the averages of the relation in Eq. (\ref{eq:inequality_action}). For the averages, we omit the dependency in $[x(t)]$ and write $\ell[x(t)]$ as a function of $t$, to shorten the notation. From the equation, the average trade-off relation is given by 
\begin{equation}
    \Delta t \ge \frac{\langle \ell^2(t)\rangle}{\mathcal{J}(t)},
    \label{eq:avg_sqrd_SL}
\end{equation}
with $\langle A\rangle \equiv \sum_{x}p_{x}(t) A_{x}(t)$. Remembering that the average $\langle j[x(t)]\rangle = \mathcal{J}$, the difference from previously obtained bounds such as in ref. \cite{Ito2018} lies in $\langle \ell^2\rangle$. We have the following
\begin{equation}
    \langle \ell^2 (t)\rangle = \int {\rm d }t~{\rm d}t'\left\langle\sqrt{\iota_{x}(t)\iota_{x}(t')}\right\rangle,
\end{equation}
in terms of system entropy one gets
\begin{equation}
    \langle \ell^2 (t)\rangle =  \int {\rm d}t~{\rm d}t' \left\langle\abs{\dot{s}_{x}^{{\rm s}}(t) \dot{s}^{\rm s}_{x}(t')}\right\rangle,\label{eq:ellsq}
\end{equation}
where the average inside the equation represents a correlation function of entropy production rate of the system between times $t$ and $t'$, meaning $\langle \ell^2\rangle$ is the sum of correlations of entropy at different times. 

It represents the second result of our work. Alternatively, one can consider weaker average bounds. One of such bounds, analogous to Eq.~(\ref{eq:inequality_action}), is given by
\begin{equation}
    \Delta t \ge \frac{\langle\ell (t)\rangle^2}{\mathcal{J}(t)}.
    \label{eq:SL_lower_bound}
\end{equation}
This bound follows the relation $  \langle \ell^2(t)\rangle \ge \langle \ell(t)\rangle^{2}$ because of the relation for the variance of $\ell[x(t)]$, $\langle \ell^2\rangle - \langle \ell\rangle^2 \ge 0$. With the results from Eqs.~(\ref{eq:inequality_action}), (\ref{eq:FI_speed_limits}), (\ref{eq:avg_sqrd_SL}), and (\ref{eq:SL_lower_bound}), we can test our predictions for a Brownian particle under a saturating drive force. 

\section{Example \label{sec:example}}

To confirm the results obtained for single-trajectory speed limits, we simulate a Langevin particle \footnote{Even though our formalism in Sec. \ref{sec:III} is developed for a master equation, we emphasize that both approaches are equivalent, see \textit{e.g.} chapter IX of reference  \cite{vanKampen}.} under a saturating driving force \cite{melo2024stochasticthermodynamicsfisherinformation, melo2024classicalgeometricfluctuationrelations}, and a simple example of discrete two-level system is explored in Appendix \ref{app:C}. This dynamics can be experimentally obtained by resorting to an RC circuit with Nyquist noise~\cite{vanKampen}. The model considers the Langevin equation
\begin{equation}
    \gamma \dot{x}(t) = -kx(t) + F(t) + \eta(t),
    \label{eq:Langevin_Overdamped}
\end{equation}
where $\eta(t)$ is a Gaussian white noise with $\langle \eta(t)\rangle = 0$, $\langle \eta(t)\eta(t')\rangle = 2\gamma\beta^{-1}\delta(t-t')$, with $\beta^{-1} \equiv k_{B}T$, meaning the particle is in contact with a heat bath at temperature $T$. The saturating drive force $F(t) = F_{0}(1 - e^{-t/\tau})$, with $F_{0}$ being the asymptotic value of the force and $\tau$ the characteristic time of saturation. We choose the initial distribution of the particle as an equilibrium distribution. Due to $F(t)$, the dynamics of the Langevin equation drives the particle out of equilibrium for $t > 0$. For $t \gg \tau$, the particle reaches a new equilibrium state. This example has been previously explored~ \cite{melo2024stochasticthermodynamicsfisherinformation, melo2024classicalgeometricfluctuationrelations} and represents an exactly solvable model where the SFI can be tested not only for time, but for other parameters such as $F_{0}$ and $\tau$, and in which the geometric fluctuation relations hold for asymptotic time.

\begin{figure}[ht!]
    \centering
    \includegraphics[width=1.0\linewidth]{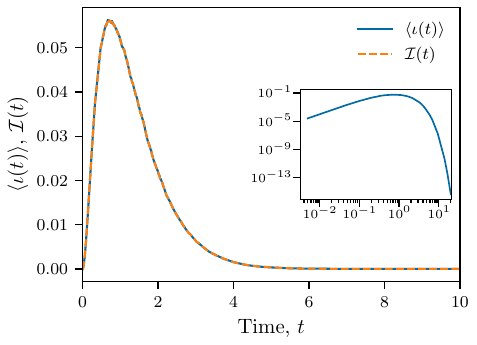}
    \caption{Comparison between the numerical average SFI and the analytical FI results for the Brownian particle under a saturating force protocol. The FI have a maxima at $t \approx \tau$, that decreases for $t > \tau$ until it reaches $\mathcal{I}(t) = 0$. The inset indicates $\langle \iota(t)\rangle$ varying with time but in a log-log scale, for better visualization.}
    \label{fig:avg_SFI_and_analytical_FI}
\end{figure}

Using stochastic path integrals, we can analytically compute the conditional probability distribution $p(x_{t},t|x_{0},0)$, and then extract the marginal probability distributions for the position at a time $t$, with $p(x,t)=\int {\rm d}x_{0}~p(x, t|x_0,0)\rho(x_0)$, where $\rho(x_0)$ is the probability distribution of the position for $t = 0$. The results for both probability distributions are detailed in Appendix \ref{app:2} together with a comparison between the analytical and simulated results of $p(x_t,t)$. Using analytical calculations of $p(x_t, t)$, we obtain the SFI $\iota_{x}(t)$. In addition, we can obtain the analytical expression for FI by the mean of SFI, $\mathcal{I}(t) = \langle \iota_{x}(t)\rangle = \int {\rm d}x~p(x, t)\iota_{x}(t)$. For our example, the distribution $P(x,t)$ is always a Gaussian distribution, meaning that for $P(x,t) = \frac{1}{\sqrt{2\pi \sigma^{2}(t)}}e^{-\frac{(x - \mu(t))^2}{2\sigma^{2}(t)}}$, we have $\mathcal{I}(t) = \left(\frac{\sigma'(t)}{\sigma(t)}\right)^2 + \left(\frac{\mu'(t)}{\sigma(t)}\right)^{2}$.

Moreover, the expression for $\iota_{x}(t)$ is analytically obtainable for a gaussian probability distribution, as $\mathcal{I}(t)$. For a gaussian distribution with mean $\mu(t)$ and variance $\sigma^{2}(t)$, it reads 
\begin{align}
 \iota_{x}(t) &= \left(\frac{\sigma'(t)}{\sigma(t)}\right)^2 + \left(\frac{\sigma'(t)(x- \mu(t))^2}{\sigma^3}\right)^2 \nonumber\\& + \left(\frac{\mu'(t)(x- \mu(t))}{\sigma^2}\right)^2 -\frac{2\mu'(t)\sigma'(t)(x- \mu(t))}{\sigma^3}\nonumber\\&- \frac{2\sigma'(t)^2(x- \mu(t))}{\sigma^4} - \frac{2\sigma'(t)\mu'(t)(x- \mu(t))^3}{\sigma^5}.~
\end{align}

We use the Euler-Maruyama method to calculate the simulations of the dynamics of Eq.~(\ref{eq:Langevin_Overdamped}). To test the simulation results, we show in Fig.~\ref{fig:avg_SFI_and_analytical_FI} that the mean SFI derived by the average SFI for all trajectories in all time steps is equivalent to computing the analytical FI for the numerical values of the average velocities. As expected, for a distribution where $\sigma(t) = \sigma = 1/\sqrt{k\beta}$, $\mathcal{I}(t) = (\mu'(t)/\sigma)^2$, meaning $\mathcal{I}(t) = k\beta\langle x'(t)\rangle^2$.

\begin{figure}[ht!]
    \centering
    \includegraphics[width=1.0\linewidth]{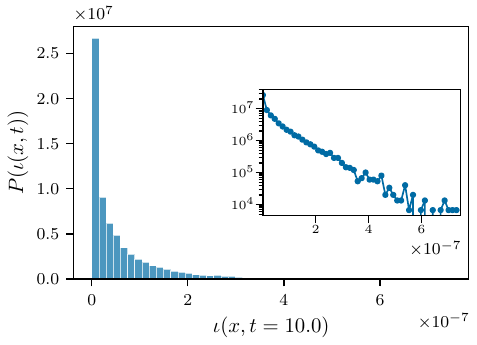}
    \caption{Probability distribution functions for the SFI at time $t = 10$ for the Brownian particle under a saturating force protocol. The results show that the SFI distribution is skewed and strictly positive. The inset shows the distribution in log-linear scale.}
    \label{fig:PDFs_for_SFI}
\end{figure}

With the values of $\iota_{x}(t)$ for all time steps and all trajectories it becomes possible to calculate the probability distributions $p(\iota_{x}(t))$ for $\iota_{x}(t)$ at a $t$ time instant. Fig.~ \ref{fig:PDFs_for_SFI} shows the PDF for the SFI at time $t = 10.0$. As expected, the distribution only admits positive values of $\iota_{x}(t)$ as it represents a quadratic quantity. 

\begin{figure}[ht!]
    \centering
    \includegraphics[width=1.0\linewidth]{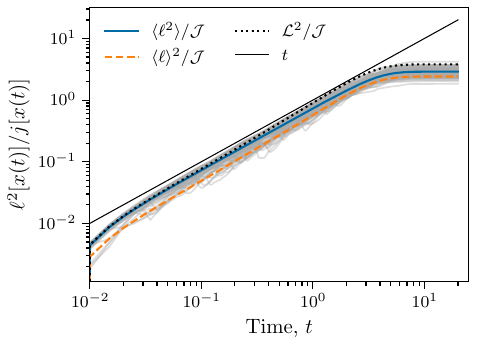}
    \caption{Single-trajectory and averaged speed limits for the Brownian particle under a saturating force protocol. The grey lines correspond to the single-trajectory speed limits, the blue line to Eq.~(\ref{eq:avg_sqrd_SL}), the orange line to Eq.~(\ref{eq:SL_lower_bound}), the black dotted line to Eq.~(\ref{eq:FI_speed_limits}), and the black straight line to the time.  Our results show that the speed limits hold on trajectory level for all trajectories in our simulations. Moreover, there is a hierarchy relation between average speed limits, as indicated in Eq.~(\ref{eq:hierarchy_SLs}). For $t < 1$, $\mathcal{L}^2 \approx \langle\ell^2\rangle$, saturating the inequality relation between both. For $t \approx 1$, the bound for $\mathcal{L}^2$ saturates to the limit, what is undone for larger values of time.}
    \label{fig:speed_limits_trajectory_level}
\end{figure}

Having the results for $\iota_{x}(t)$ and $\mathcal{I}(t)$, it becomes possible to compute the speed limits in trajectory level, and the average speed limits for Eqs.~(\ref{eq:inequality_action}),~(\ref{eq:FI_speed_limits}),  (\ref{eq:avg_sqrd_SL}), and \ref{eq:SL_lower_bound}), respectively. It is possible to propose a hierarchy relation of average speed limits 
\begin{equation}
    \tau \ge \frac{\mathcal{L}^2(t)}{\mathcal{J}(t)} \ge \frac{\sigma_{\ell}^{2}(t)}{\mathcal{J}(t)} \ge 0,
    \label{eq:hierarchy_SLs}
\end{equation}
where $\sigma_{\ell}^{2}(t) \equiv \langle \ell^2(t)\rangle - \langle\ell(t)\rangle^2$. Eq.~(\ref{eq:hierarchy_SLs}) is observed in Fig. \ref{fig:speed_limits_trajectory_level}. The black straight line represents $t$, the black dotted line represents Eq.~(\ref{eq:FI_speed_limits}), the blue line represents Eq.~(\ref{eq:avg_sqrd_SL}), and the orange line represents Eq.~(\ref{eq:SL_lower_bound}). The gray lines represent Eq.~(\ref{eq:inequality_action}), confirming our predictions for speed limits on trajectory level. Our results show that the hierarchy is respected during the process, and that only Eq.~(\ref{eq:FI_speed_limits}) saturates the bound at $t \approx 1$, which is the value of the characteristic time for the process. Furthermore, the bound between Eqs.~(\ref{eq:FI_speed_limits}) and (\ref{eq:avg_sqrd_SL}) saturates for $0 < t < 1$. The speed limits in trajectory level only obey the inequality of Eq.~(\ref{eq:inequality_action}), with some trajectories having limits tighter than $\mathcal{L}^2(t)/\mathcal{J}(t)$ apart from the region it saturates.

An alternative, physically equivalent limit to the overdamped Brownian system treated above is the use of a  drive force $F(t) = F_{0}e^{-t/\tau}$, which represents a change of variables of $x(t) \rightarrow -x(t) + F_0/k$. Even though this represents a direct mapping between both Langevin equations, the qualitative and quantitative results for the SFI differ. The model is described by a Langevin equation much like Eq.~(\ref{eq:Langevin_Overdamped}), with $F(t)$ as described above. The driving force is $F_{0}$ at the beginning, and decays exponentially as time elapses until it reaches $\lim_{t\rightarrow \infty}F(t) = 0$. Using stochastic path integrals, we also obtain the conditional probability $P(x, t|x_{0},0)$, and obtain the marginal probability of the particle being a position $x$ at a time $t$ via $P(x,t) = \int {\rm d}x_{0}~P(x,t|x_{0},0)\rho(x_{0})$, with $\rho(x_{0})$ being an equilibrium distribution. The resulting $P(x,t)$ distribution is also a Gaussian distribution, with mean 
\begin{align}
    \langle x (t)\rangle &= \frac{e^{-t\left(\frac{k}{\gamma} + \frac{1}{\tau}\right)}F_{0}\tau\sqrt{e^{\frac{2kt}{\gamma}} + e^{\frac{2t}{\tau}} - 2e^{t\left(\frac{k}{\gamma} + \frac{t}{\tau}\right)}}}{\sqrt{(\gamma  - k\tau)^2}},\\
    \sigma^2(t) &= \frac{1}{k\beta}.
\end{align}
On average, the particle reaches a maximum and goes back to origin, where it achieves equilibrium.

\begin{figure}[ht!]
    \centering
    \includegraphics[width=1.0\linewidth]{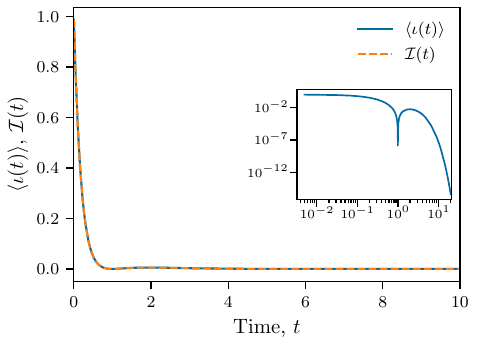}
    \caption{Comparison between the numerical average SFI and the analytical FI results for the Brownian equation under an evanescing force protocol. The FI have a maxima at $t \approx \tau$, that decreases for $t > \tau$ until it reaches $\mathcal{I}(t) = 0$. The inset indicates $\langle \iota(t)\rangle$ varying with time but in a log-log scale, for better visualization.}
    \label{fig:avg_SFI_and_analytical_FI_other}
\end{figure}

Again, we use the Euler-Maruyama method to calculate the simulations of the dynamics. To test the simulation results, we show in Fig.~\ref{fig:avg_SFI_and_analytical_FI_other} that the mean SFI derived by the average SFI for all trajectories in all time steps is equivalent to computing the analytical FI for the numerical values of the average velocities, with $\mathcal{I}(t) = k\beta\langle x'(t)\rangle^2$. 

\begin{figure}[ht!]
    \centering
    \includegraphics[width=1.0\linewidth]{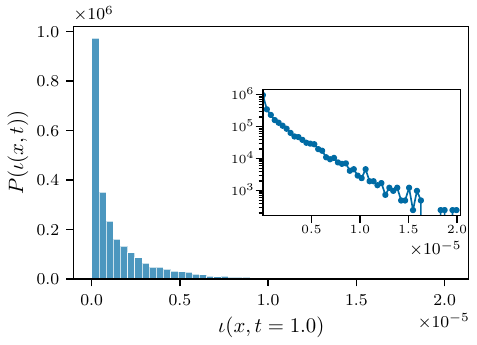}
    \caption{Probability distribution functions for the SFI at time $t = 10$, for the Brownian equation under an evanescing force protocol. The results show that the SFI distribution is skewed and strictly positive. The inset shows the distribution in log-linear scale.}
    \label{fig:PDFs_for_SFI_other}
\end{figure}

With the values of $\iota_{x}(t)$ for all time steps and all trajectories it becomes possible to calculate the probability distributions $P(\iota_{x}(t))$ for $\iota_{x}(t)$ at time $t$. Figure\ref{fig:PDFs_for_SFI_other} shows the PDF for the SFI at time $t = 10.0$. As expected, the distribution only admits positive values of $\iota_{x}(t)$ as it represents a quadratic quantity. Also, the distribution is centered close to $\iota(x,t) = 0$. 

\begin{figure}
    \centering
    \includegraphics[width=1.0\linewidth]{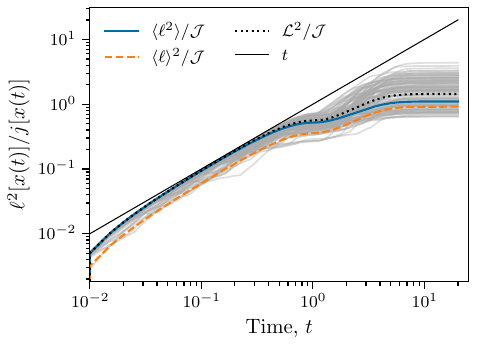}
    \caption{Single-trajectory and averaged speed limits for the Brownian equation under an evanescing force protocol. The gray lines correspond to the single-trajectory speed limits, the blue line to Eq.~(\ref{eq:avg_sqrd_SL}), the orange line to Eq.~(\ref{eq:SL_lower_bound}), the black dotted line to Eq.~(\ref{eq:FI_speed_limits}), and the black straight line to the time.  Our results show that the speed limits hold on trajectory level for all trajectories in our simulations. The qualitative behavior of single-trajectory speed limits resembles the behavior of the average speed limits. Moreover, As the system is driven to equilibrium, the limits tend to a constant.}
    \label{fig:speed_limits_trajectory_level_other}
\end{figure}
We can also plot the results for the speed limits from Eqs.~(\ref{eq:inequality_action}) and Eq.~(\ref{eq:hierarchy_SLs}). The results are synthesized in Fig.~\ref{fig:speed_limits_trajectory_level_other}. The grey lines indicate the single-trajectory speed limits, the black dotted line the speed limit for the FI, Eq.~(\ref{eq:FI_speed_limits}), the blue line indicates the upper bound for the average speed limit for the SFI, Eq.~(\ref{eq:avg_sqrd_SL}), and the orange dashed line indicates the lower bound for the average speed limit for the SFI, Eq.~(\ref{eq:SL_lower_bound}). Similar to results plotted in Fig.~\ref{fig:speed_limits_trajectory_level}, the speed limits (averaged and single-trajectory) reflect the relaxation process to equilibrium of the system. This happens because the speed limit is proportional to the length and the irreversible work. 
\section{Discussions and Conclusions}
In this paper, we have delved into two questions regarding the SFI framework.~Firstly, we have introduced a stochastic thermodynamic interpretation for the stochastic Fisher information and for quantities originated from it, such as the stochastic length and the stochastic action in the context of systems of discrete energy levels. Finally, we set forth the notion of single-trajectory speed limits that emerges from the geometric analysis on trajectory level. By averaging the inequalities one has obtained a hierarchy of speed limits, that is observed to happen in a system of Brownian particle under a saturating drive force. The single-trajectory speed limits are also observed to be respected for all trajectories, with some fluctuations saturating the bound for the region around the maxima of $\mathcal{I}(t)$. 

For the interpretation of the SFI, we have found that contrary to the stochastic thermodynamic interpretation of the FI, it is an analogous interpretation for the stochastic line element ${\rm d}s^2(x,t)/{\rm d}t^2$, except that is not given in terms of the expected rate weighted value for the rate of entropy production in the environment ${\rm d}s_{x\rightarrow x'}^{{\rm e}}(t)/{\rm d}t$ and the thermodynamic force rate ${\rm d}F_{x\rightarrow x'}(t)/{\rm d}t$, but by the rate-weighted expected value of the rate of production of entropy of the system times the exponential of the thermodynamic force. For equilibrium distributions, the stochastic line element is given by contributions to the variance of the rate of change of energy. As expected, this result show that the SFI is the stochastic equivalent to the FI. 

For single-trajectory speed limits, our results demonstrate that the Stochastic Fisher Information (SFI) metric enables the derivation of speed limits at the trajectory level. Furthermore, averaging these trajectory-specific bounds yields new, less restrictive speed limits compared to those obtained via conventional Fisher Information (FI). This systematic approach establishes a hierarchy of speed limits. We also proposed a physical intuition behind the single-trajectory speed limits. Considering the previous results for stochastic speed limits, where the entropic velocity of transformation between two states induced by a protocol $\lambda(t)$ is upper bounded by the irreversible entropy production during the realization of such protocol, we propose that the single-trajectory speed limits have an analogous behavior, where given a protocol $\lambda(t)$ the stochastic entropic velocity for the transformation of a system between two states is upper bounded by the stochastic irreversible entropy production. 

To validate this framework, we applied it to two examples of Brownian particle subjected to driving forces (protocols). The first one was for a saturating force, and the second for a decreasing force. The results confirm that speed limits hold both at the single-trajectory level and on average. Specifically, single-trajectory speed limits saturate in the regime where $\mathcal{L}^2/\mathcal{J}$ saturates, while in non-saturating regimes, certain individual trajectories exhibit tighter bounds than any corresponding averaged relations. Interestingly, for $t < 1$, $\mathcal{L}^2 \approx \langle \ell^2\rangle$ and $\langle \ell^2\rangle > \langle \ell^2\rangle$, but there is a transition on the region of $t \approx 1$ and for $t > 1$, $\mathcal{L}^2 > \langle \ell^2\rangle$, and $\langle \ell^2\rangle \rightarrow \langle \ell\rangle^2$ as $t$ tends to infinity.

To expand physical insight about single-trajectory speed limits, consider a regime of near-equilibrium, with the canonical probability distribution $p(x,t) = e^{-\beta\varepsilon(x,t)}/Z(\beta,t)$. Eq.~(\ref{eq:inequality_action}) reads
\begin{widetext}
\begin{equation}
    \Delta t \ge \frac{\ell^2[x(t)]}{j[x(t)]} = \frac{\iint {\rm d}t{\rm d}t' \sqrt{\left(\pdv{\varepsilon(x,t)}{t} -  \left\langle \pdv{\varepsilon(x,t)}{t}\right\rangle\right)^2 \left(\pdv{\varepsilon(x,t')}{t'} -  \left\langle \pdv{\varepsilon(x,t')}{t'}\right\rangle\right)^2}}{\int {\rm d}t\left(\pdv{\varepsilon(x,t)}{t} -  \left\langle \pdv{\varepsilon(x,t)}{t}\right\rangle\right)^2}, \label{eq:SSLs_Canonical}
\end{equation}
\end{widetext}
which resembles an autocorrelation function, without the average. Equation~(\ref{eq:SSLs_Canonical}) is reminiscent of previous results in the literature, namely those of Refs.~\cite{crooks2007, Sivak2012}. As proposed, we observed an identical qualitative behavior on trajectory level, indicating that the interpretation for the single-trajectory speed limit also follows such reasoning, with the restriction being a stochastic irreversible work associated to a single trajectory.  

For outlook, many questions arise. The introduction of concepts such as a formal geometric definition of stochastic information geometry remains beyond the scope of this paper. Other questions include when to expect the saturation of the bounds of Eq.~(\ref{eq:hierarchy_SLs}), for which systems the limits become more restrictive, and the relation between the single-trajectory speed limits and thermodynamic quantities such as work and heat. The present work opens an avenue for the further application of the SFI towards a physical understanding of (non-equilibrium) systems. The existence of speed limits at the trajectory level is likely to prove itself very useful in experiments, where one has to deal with apparatus uncertainty and natural fluctuations induced by the environment. Moreover, while our derivations are made for overdamped dynamics, our formalism holds to underdamped and non-markovian dynamics, with due changes. As a perspective, we are working to generalize the SFI formalism to underdamped Brownian systems and non-markovian systems, to look for the effects of kinect energy terms on the SFI and on trajectory level speed limits, as well as generalization of previous results such as the classical geometric fluctuation relations for the SFI \cite{Melo_2025}.

During the elaboration of this work, we came across to a work with conceptual overlaps to our results \cite{lyu2025learningstochasticthermodynamicsdirectly}.

\section*{Acknowledgments} 
PBM acknowledges Alice M. Aredes and Pedro V. Paraguass\'u for enlightening discussions, and Pedro E. Harunari for raising interesting questions. 

This work is supported by the Brazilian agencies Conselho Nacional de Desenvolvimento Científico e Tecnológico (CNPq), Coordena\c{c}\~{a}o de Aperfei\c{c}oamento de Pessoal de Ensino Superior (CAPES), and Funda\c{c}\~{a}o Carlos Chagas de Apoio \`a Pesquisa do Estado do Rio de Janeiro (FAPERJ). PBM acknowledges CAPES' scholarship, finance code 001. FI acknowledges financial support from CAPES, CNPQ, and FAPERJ (Grants No. 308205/2019-7, No. E-26/211.318/2019, No. 151064/2022-9, and No. E-26/201.365/2022) and by the Serrapilheira Institute (Grant No. Serra 2211-42166). DOSP acknowledges the Brazilian funding agencies CNPq (Grants No. 307028/2019-4, 402074/2023-8), FAPESP (Grant No. 2017/03727-0), and the Brazilian National Institute of Science and Technology of
Quantum Information (INCT-IQ) Grant No. 465469/2014-0. SMDQ acknowledges CNPq grant No. 302348/2022-0, and FAPERJ grant No. APQ1-210.310/2024. WAMM acknowledges CNPq grant No. 308560/2022-1.
\appendix

\section{Derivation of Eqs.~(\ref{eq:19}) -- (\ref{eq:21}) \label{app:1}}

To appreciate the results of Sec.~\ref{sec:III}, we detail the calculations for the stochastic thermodynamic interpretation for Eqs.~(\ref{eq:19})--(\ref{eq:28}). The definition of SFI is given by
\begin{eqnarray}
    \frac{{\rm d}s^2(x,t)}{{\rm d}t^2} = \left(\dv{\log p_{x}(t)}{t}\right)^2,
\end{eqnarray}
and by expansion it can be written as
\begin{align}
    \dv{s^2(x,t)}{t^2} &=  \frac{1}{p_{x}(t)}\dv{t}\left(\dv{p_{x}(t)}{t}\right)\nonumber - \dv{t} \left( \dv{\log p_{x}(t)}{t}\right)\\
    &= (1) + (2).
\end{align}

By expressing the first and the second terms in terms of the master equation, one gets
\begin{align}
    (1) &= \frac{1}{p_{x}(t)} \dv{t}\left(\sum_{x'}W_{x\rightarrow x'}(t) ~p_{x'}(t)\right),\\
    &= \frac{1}{p_{x}(t)} \dv{t}\left(\sum_{x'}W_{x\rightarrow x'}(t) e^{-F_{x\rightarrow x'}(t)}p_{x}(t)\right),\\ &= \sum_{x'}\left(\dv{W_{x\rightarrow x'}}{t}e^{-F_{x\rightarrow x'}} - W_{x\rightarrow x'}e^{-F_{x\rightarrow x'}} \dv{F_{x\rightarrow x'}}{t} \nonumber\right.\\&\left.+ W_{x\rightarrow x'}e^{-F_{x\rightarrow x'}}\dv{\log p_{x}}{t}\right),~~
\end{align}
where we omit the $t$ dependence for simplicity.

The second term gives 
\begin{align}
    (2) &= - \dv{t} \left(\frac{1}{p_{x}(t)} \dv{p_{x}(t)}{t}\right),\\
    &= -\dv{t}\left(\sum_{x'}W_{x\rightarrow x'}e^{-F_{x\rightarrow x'}}\right),\\
    &= -\sum_{x'}\left(\dv{W_{x\rightarrow x'}}{t}e^{-F_{x\rightarrow x'}} \right.\nonumber\\&\left.- W_{x\rightarrow x'}e^{-F_{x\rightarrow x'}} \dv{F_{x\rightarrow x'}}{t}\right).
\end{align}

By summing $(1) + (2)$ one gets
\begin{align}
    (1) + (2) &= \left(\sum_{x'} W_{x\rightarrow x'}(t)e^{-F_{x\rightarrow x'}}\dv{\log p_{x}(t)}{t}\right).\label{eq:A9}
\end{align}
The equation above represents a description of the SFI in terms of master equation related quantities. The time derivative of $\log p_x(t)$  cannot be connected to a thermodynamic quantity such as the stochastic entropy production rate of the system $\dot{s}^{{\rm s}}(x(t),t)$ without first establishing the limit $x \rightarrow x(t)$. The reason behind such a transformation is that the SFI is a local quantity in time, which means that the position $x$ represents the marginal probability distribution in space when observed from the perspective of the probability distribution associated with the whole trajectory.

To associate Eq.~(\ref{eq:A9}) to an entropy-related factor, one needs to take the following limit $x \rightarrow x(t)$, and therefore Eq.~(\ref{eq:A9}) becomes
\begin{align}
    \left.\frac{{\rm d}s^2(x,t)}{{\rm d}t^2}\right|_{x(t)} = \left(\sum_{x'} W_{x\rightarrow x'}(t)e^{-F_{x\rightarrow x'}}\left.\pdv{\log p_{x}(t)}{t}\right|_{x(t)}\right),
\end{align}
where $\pdv{\log p_{x(t)}(t)}{t} = -\dot{s}_{x(t)}^{{\rm s}}(t) + \sum_{j}\delta(t - t_j)\log(p_{x_j^+}(t_j)/p_{x_j^-}(t_j))$, with the $\delta$ term meaning the entropy contribution to the jumps between levels along a trajectory. The equation above can be rewritten as
\begin{align}
\frac{\partial_t p_{x(t)}(t)}{p_{x(t)}(t)} &= -\dot{s}_{tot}(t) - \sum_{j} \delta(t - t_j)\log \left(\frac{p_{x_j^+}W_{x_j^+ \rightarrow x_{j}^{-}}}{p_{x_j^-}W_{x_j^- \rightarrow x_{j}^{+}}}\right),\\
&= -\dot{s}_{tot}(t) + \sum_{j} \delta(t - t_j)F_{x_{j}^{-}\rightarrow x_{j}^{+}}(t_j),
\end{align}
where in first expression we suppressed the time-dependency on the second term in right to shorten notation, and we also suppressed $\dot{s}_{tot}(t)$ trajectory dependency. Notice that Eq.~(A11) still represents the time rate of $\log p_{x(t)}(t)$. 

We can now write an appropriate thermodynamic expression for the SFI, being
\begin{align}
    \left.\frac{{\rm d}s^2(x,t)}{{\rm d}t^2}\right|_{x(t)} &= \left(\sum_{x'} W_{x\rightarrow x'}(t)e^{-F_{x\rightarrow x'}}\left.\pdv{\log p_{x}(t)}{t}\right|_{x(t)}\right),\\
    \left.\frac{{\rm d}s^2(x,t)}{{\rm d}t^2}\right|_{x(t)}&= \left(\overline{e^{-F_{x\rightarrow x'}}\left(- \dot{s}_{tot}(t)\right.} \right.\nonumber\\+& \left.\overline{\left.\sum_{j}\delta(t - t_j)F_{x^{-}_j\rightarrow x^{+}_j}(t_j)\right)} \right),
\end{align}
where $\overline{A} \equiv \sum_{x'}W_{x\rightarrow x'}(t) A_{x'\rightarrow x}(t)$ represents the rate-weighted expected value of $A_{x'\rightarrow x}$.

\section{Full expressions for mean and variance of the marginal probability for the adopted example \label{app:2}}

In this appendix, we show the expressions for the mean and variance of $P(x,t)$ for the example in Sec. \ref{sec:example}. For the adopted overdamped Langevin equation
\begin{equation}
    \gamma \dot{x}(t) = -kx(t) + F_{0}(1 - e^{-t/\tau}) + \eta(t),
\end{equation}
by using path integral methods \cite{Wio_Path}, we calculate the transition probability $P[x_t,t|x_{0},0]$. Given an initial equilibrium distribution $\rho(x_{0})$, the marginal probability distribution $P(x_t,t)$ is a Gaussian distribution given by
\begin{equation}
    P(x,t) = \frac{1}{\sqrt{2\pi \sigma^{2}(t)}}\exp\left(-\frac{(x - \mu(t))^2}{2\sigma^{2}(t)}\right),
\end{equation}
where $\mu(t) = \langle x(t)\rangle$ is the mean value of $x$, and $\sigma^2(t) = \langle x^2\rangle(t) - \mu(t)$ is the variance of $x$. Both are given by
\begin{align}
    \langle x(t)\rangle &= \frac{F_0 \left\{\gamma\left(1 -   e^{-\frac{k t}{\gamma }}\right)+k \tau  \left(e^{-\frac{t}{\tau
   }}-1\right)\right\}}{k (\gamma -k \tau )},\\
   \sigma^{2}(t) &= \frac{1}{k\beta}.
\end{align}
\section{Example of trajectory speed limits on a discrete system} \label{app:C}

To expand our insights into trajectory level speed limits, we also can explore it in discrete systems. For instance, we consider a two-level system with rate matrix ${\bf{W}}(t)$ given by
\begin{align}
    {\bf{W}}(t) = \mqty(-W_{21}(t)&W_{12}(t)\\W_{21}(t)&-W_{12}(t)),
\end{align}
with master equations given by
\begin{align}
\dv{p_1(t)}{t} &= W_{12}(t)p_{2}(t) - W_{21}(t)p_{1}(t),\label{eq:ME_1}\\
\dv{p_2(t)}{t} &= W_{21}(t)p_{1}(t) - W_{12}(t)p_{2}(t),
\end{align}
where in principle ${\bf{W}}(t)$ is a general rate matrix. To uncouple the set of master equations we take into account the probability conservation relation $p_2(t) = 1 - p_{1}(t)$. By substituting $p_{2}(t)$ into Eq.~(\ref{eq:ME_1}) we get
\begin{align}
    \dv{p_1(t)}{t} = W_{12}(t)(1 - p_{1}(t)) - W_{21}(t)p_{1}(t),
\end{align}
where $p_{1}(t)$ gives
\begin{align}
    p_{1}(t) &= \mu^{-1}(t)\left(\int_{0}^{t}\mu(t')W_{12}(t'){\rm d}t'\right),
\end{align}
with $\mu(t) =\exp(\int_{0}^{t}(W_{12}(t') + W_{21}(t')){\rm d}t')$ and $p_{2}(t) = 1 - p_{1}(t)$.

Let us consider the simplest case, where ${\bf W}(t) = {\bf W}$ is a constant rate matrix and initial condition $p_{1}(0) = p_{2}(0) = \frac{1}{2}$. Therefore, we have
\begin{align}
    p_{1}(t) &=\frac{W_{12} \left(1-e^{-t(W_{12}+W_{21})}\right)}{W_{12}+W_{21}}+\frac{1}{2},\\
    p_{2}(t) &= \frac{1}{2} - \frac{W_{12} \left(1-e^{-t(W_{12}+W_{21})}\right)}{W_{12}+W_{21}}
\end{align}

To compute the stochastic Fisher information we have two possible results, analytically calculable, in general it means
\begin{align}
\iota_{i}(t) = \left(\dv{\log p_{i}(t)}{t}\right)^2,\end{align}
and the stochastic length $\ell[x(t)]$ becomes
\begin{align}
    \ell[x(t)] = \int {\rm d}t \sqrt{\iota_{x[t]}(t)}.
\end{align}
For each state, $\iota_{i}(t)$ is given by
\begin{align}
    \iota_{1}(t) &= \frac{4 W_{12}^2 \left(W_{12}+W_{21}\right)^2}{\left(\left(3 W_{12}+W_{21}\right) e^{t\left(W_{12}+W_{21}\right)}-2 W_{12}\right)^2},\\
    \iota_{2}(t) &= \frac{4 W_{12}^2 \left(W_{12}+W_{21}\right)^2}{\left(\left(W_{12}-W_{21}\right) e^{t\left(W_{12}+W_{21}\right)}-2 W_{12}\right)^2}.
\end{align}
Discretizing time and taking time steps of size $\Delta t$, one has that $\ell[x(t)]$ is approximately
\begin{align}
    \ell[x(t)] \approx \sum_{k = 1}^{N}\sqrt{\iota_{i_k}(t_k))}\Delta t,
\end{align}
which for a two-level system with $\Delta t = 1$, considering that the system spends $n_1$ times on state $1$ and $n_2 = N-n_1$ times on state $2$, is given by
\begin{align}
    \ell[x(t)] \approx \sum_{k=1}^{n_1}\sqrt{\iota_{1}(t_k)} + \sum_{k=1}^{n_2}\sqrt{\iota_{2}(t_k)}.
\end{align}

To understand the trajectory-level speed limits on has that, from Cauchy-Schwarz theorem,
\begin{align}
    \left(\int {\rm d}t \sqrt{\iota_{x[t]}(t)}\right)^2 &\le \iint{\rm d}t''{\rm d}t'~\iota_{x[t]}(t'),\\
    \left(\int {\rm d}t \sqrt{\iota_{x[t]}(t)}\right)^2 &\le \Delta t \int{\rm d}t'~\iota_{x[t]}(t'),\\
    \Delta t &\ge \frac{(\ell[x(t)])^2}{j[x(t)]},
\end{align}
which on a discrete-time approximation it results into
\begin{align}
    \frac{\ell^2}{j} &\approx \frac{\sum_{k,k'}(\Delta t)^2 \sqrt{\iota_{i_k}(t_k)\iota_{i_{k'}}(t_{k'})}}{\Delta t\sum_{k} (\Delta t)\iota_{i_k}(t_k)},
\end{align}
where $\sqrt{\iota_{i_k}(t_k)\iota_{i_{k'}}(t_{k'})} \le \iota_{i_k}(t_k)$ for all $\{t_k\}$ and $\{t_{k'}\}$.

It follows that
\begin{align}
   \frac{\sum_{k,k'}(\Delta t)^2 \sqrt{\iota_{i_k}(t_k)\iota_{i_{k'}}(t_{k'})}}{\Delta t\sum_{k} (\Delta t)\iota_{i_k}(t_k)} &\le 1,\\
   \frac{\sum_{k,k'}(\Delta t)^2 \sqrt{\iota_{i_k}(t_k)\iota_{i_{k'}}(t_{k'})}}{\sum_{k} (\Delta t)\iota_{i_k}(t_k)} &\le \Delta t.   
\end{align}
One could extend such derivation for more complex settings, such as discrete systems with more levels and for time-dependent ${\bf{W}}(t)$ matrices. The relation still holds.

\providecommand{\newblock}{}
\bibliography{refs}




\end{document}